\documentclass[%
 reprint,
 amsmath,amssymb,
 prl
]{revtex4-2}

\usepackage{graphicx}
\usepackage{dcolumn}
\usepackage{bm}
\usepackage[mathlines]{lineno}
\modulolinenumbers[5]

\usepackage{upgreek,color}
\newcommand{\dspl}[1]{\displaystyle{#1}}

\newcommand{\demi}{\frac{1}{2}}
\newcommand{\mb}[1]{\mbox{\boldmath$#1$}}
\newcommand{\mbt}{\mb{t}}
\newcommand{\mbk}{\mb{k}}
\newcommand{\mbr}{\mb{r}}
\newcommand{\umb}[1]{\hat{\mbox{\boldmath$#1$}}}
\newcommand{\umbk}{\umb{k}}

\newcommand{\umbs}{\umb{s}}

\newcommand{\umbu}{\umb{u}}
\newcommand{\umbv}{\umb{v}}

\newcommand{\umbn}{\umb{n}}
\newcommand{\Curv}[1]{\mathbb{#1}}

\begin{document}

\preprint{APS/123-QED}

\title{Ray Theory of Waves}
\author{Kuan Fang Ren$^{1*}$, Qingwei Duan$^1$, Claude Rozé$^2$, Minglin Yang$^{3*}$, Ce Zhang$^{4}$, Haiping Fang$^5$, and Xiang'e Han$^{1}$ }
\email{Corresponding authors: \\ren.kuanfang@orange.fr, \\xehan@mail.xidian.edu.cn, \\yangminglin@bit.edu.cn}
\affiliation{%
 $^1$School of Physics, Xidian University, Xi’an 710071, China
\\
$^2$CORIA/UMR 6614 CNRS - University of Rouen Normandy,
76801 Saint-Etienne du Rouvray, France
\\
$^3$Institute of Radio Frequency Technology and Software, School of Integrated Circuits and Electronics, Beijing Institute of Technology (BIT), Beijing 100081, China
\\
$^4$School of Information Science and Engineering, Southeast University, Nanjing 210096, China
\\
$^5$School of Physics, East China University of Science and Technology, Shanghai, 200237, China}%

\date{\today}
             
\begin{abstract}
Accurate and efficient prediction of three-dimensional (3D) fields in wave interactions with large, complex-shaped objects is essential for applications in electromagnetic computation, computer graphics, optical metrology, and freeform optics. However, existing methods face significant challenges: numerical techniques are computationally intensive and impractical for large objects, while ray tracing neglects wave properties and remains inefficient, relying solely on ray bundles. In this Letter, we present the \emph{Ray Theory of Waves} (RTW), which introduces \emph{wavefront curvature (WFC) as an intrinsic property of a ray} to describe wave divergence and convergence. Using differential geometry, we derive the wavefront equation, rigorously relating WFC of incident, reflected, and refracted waves, enabling accurate calculation of field amplitude and phase along a ray. To address diffraction effects at singularities and compute the total field, we propose an anti-conventional strategy. The flexibility, precision and performance of RTW are demonstrated through the calculation of 3D scattering pattern of an ellipsoidal drop. Importantly, the method clarifies several longstanding queries about Airy theory since the 19th century. RTW constitutes a theoretical breakthrough, opening new avenues for practical applications.
\end{abstract}

\maketitle


\textbf{Introduction} -- The interaction of waves (light, electromagnetic or acoustic waves) with large macroscopic objects is of great importance in numerous scientific and engineering domains, including light scattering by particles\cite{vandeHulst57,KharePRL1974,NyeNature1984,Berry80},
computational electromagnetics (antennas, radar, wireless ...)\cite{Taflove2015,Yang2019,NotarosIEEE-APM75years}, object visualization in computer graphics\cite{GlassnerRayTr1989,PharrBook2004,RayTrGemsII2021}, optical measurement techniques\cite{HollPRL2017,PanNature2024}, and design of freeform optics lenses\cite{RollandOptica2021,YangOE2021}. 
A critical challenge in these fields is the accurate and efficient prediction of three-dimensional (3D) field in wave interaction with complex-shaped objects that are much larger than the wavelength. Despite significant progress, existing methods remain insufficient for scientific and engineering needs. 
Numerical techniques have been extensively developed for electromagnetic (EM) computation \cite{ShengSong2012,WuYM2016,SongIEEE1997} and light scattering\cite{Mishchenko00,YurkinJQSRT2007DDA}. Among them, the multilevel fast multipole algorithm (MLFMA) \cite{SongIEEE1997,Yang2019} is particularly powerful. However, these methods are computationally expensive and unsuitable for large objects, as they are limited to sizes (typically below 200 wavelengths for dielectric materials)\cite{Yang2019}. While rigorous analytical solutions exist for objects with simple geometries (e.g., spheroids, ellipsoids), they suffer from size limitations comparable to numerical approaches. Various approximate methods have also been developed to address specific problems\cite{Marston94,Mishchenko00,BiYang2013,WuYM2016,AdamBook2017,Guo2022IEEE}. For instance, the catastrophe theory provides a powerful tool for analyzing the morphology of the scattering patterns near caustics using polynomial phase functions\cite{Berry80,AdamBook2017}, yet it cannot predict full scattering pattern of \emph{real} objects. Thus, there is an urgent need for a method that can simulate \emph{directly} and \emph{accurately} 3D field in wave interaction with large, complex-shaped objects.

The ray model offers a promising approach to this challenge due to its inherent flexibility. However, it fails to account for wave properties, particularly wavefront shape. The field intensity variation is then calculated using energy tube in EM computation, ray bundles in optics and statistical technique of ray numbers in computer graphics.  Yet, all these methods remain approximate, requiring a large number of rays to achieve acceptable precision. Moreover, in EM computation, phase plays an important role, but the classical ray model struggles to calculate the phase of focal lines. 

In this Letter we develop the \emph{Ray Theory of Waves} (RTW) based on \emph{a novel ray concept}: rays representing a wave possess inherently \emph{wavefront curvature} (WFC). The core principles of RTW are as follows (i).  All EM waves are described by rays characterized by the wave vector $\mbk$, electric vector $\mb{E}$ and WFC matrix $\Curv{Q}$. (ii). $\mb{E}$ at any point along a ray is rigorously calculated using Vectorial Complex Ray Model (VCRM). (iii). In regular regions, the total field is the sum of the complex amplitudes of all the rays at that point. In the vicinity of singular points, singularity theory\cite{Berry2023,Berry1969,stamnes86} is applied to account diffraction effect and compute the total field, adopting an anti-conventional strategy.

The novel property WFC of the rays enables describing the wave shape and allows for \emph{rigorous} calculation of both amplitude and phase at any point along a ray. The relationship between WFC of incident, refracted, and reflected waves is governed by the \emph{wavefront equation}, which we derive using differential geometry, leading to the establishment of VCRM. The flexibility, accuracy, and performance of VCRM are demonstrated by comparing the 3D scattering pattern of an ellipsoidal drop with results from MLFMA. As a simple yet significant example, we show that RTW provides a rigorous framework with a clear physical explanation for longstanding questions regarding Airy theory of the rainbow in spherical drops since the 19th century. While previous approximate methods\cite{Konnen:79,KharePRL1974} improved solutions, they did not fully reveal the underlying physical mechanism. The direct application of the same procedure to a non-spherical drop confirms applicability and performance of RTW. 

\textbf{Vectorial Complex Ray Model in a symmetric plane} --
When the incident plane coincides with a \emph{symmetric plane} of the object, all reflected and refracted rays are confined to this plane and the two principal curvatures of the wavefronts and the object surface remain independent. Using Hamilton's method and the concept of a thin bundle of rays, Synge (1937) derived the relationship between two principal WFC radii of the incident wave and those of the refracted or reflected wave\cite{Synge1937}. These relations have been applied in the study of image aberration\cite{born99}. However, their application to EM/light scattering has not been reported, nor has any attempt been made to extend Hamilton's method to 3D wave-object interaction.
\vspace{-3mm}
\begin{figure}[!ht]
	\begin{center}
		\includegraphics[width=0.3\textwidth]{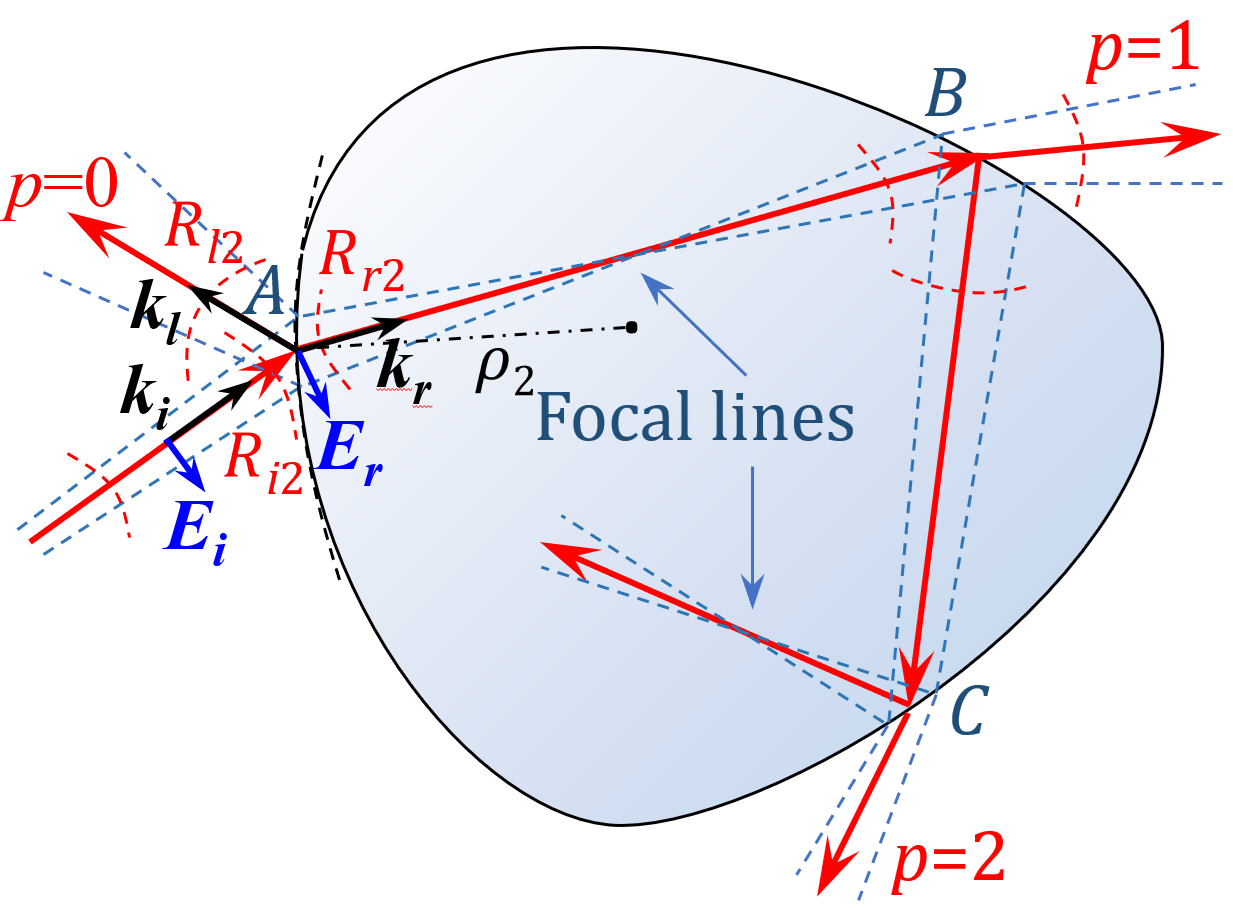}
		\caption{Ray tracing in VCRM-Sym. $R_{x\mu}$ represents WFC radius and  $\rho_\mu$ denotes the curvature radius of object surface. The index $x=i, l$ or $r$ corresponds to the incident, reflected or refracted wave, respectively, while $\mu=1$ or 2 indicates the direction perpendicular to or within the incident plane. $\mb{k}$ is the wave vector and $p$ the order of rays\cite{vandehulst80m}.}
		\label{fig:RayTracing}
	\end{center}
\end{figure}\vspace{-6mm}

By applying the novel concept of rays and differential geometry, we derived Synge's relations for a refracted ray, given as follows:
\begin{eqnarray}
	\label{eq:WavefrontEqSym1a}
	\frac{k_r}{R_{r1}} &=& \frac{k_i}{R_{i1}}+\frac{k_{rn}-k_{in}}{\rho_1}\\
	\label{eq:WavefrontEqSym2a} 
	\frac{k_{rn}^2}{k_rR_{r2}} &=& \frac{k_{in}^2}{k_iR_{i2}} +\frac{k_{rn}-k_{in}}{\rho_2}
\end{eqnarray}
where $k$ denotes the wave number, the index $_n$ indicates the normal component. The meanings of the other symbols are explained in the caption of Fig. \ref{fig:RayTracing}. Based on this, we have developed VCRM for scattering in a symmetric plane\cite{RenVCRMOptLett2011} (VCRM-Sym). Fig. \ref{fig:RayTracing} compares ray tracing in VCRM-Sym with the classical ray model. In the symmetric plane, the classical ray model requires two rays (dashed lines) to approximate the field intensity, whereas VCRM-Sym achieves a rigorous calculation using a single ray (red line). Moreover, the phase of focal lines can be readily determined by simply counting their occurrences.

The results of VCRM-Sym have been validated against experimental measurements\cite{Fabrice15OE} and compared to those of MLFMA, demonstrating that VCRM-Sym is $10^4$ times faster\cite{YangCompJQSRT15}. The precision is excellent except near caustics. 

\textbf{Vectorial Complex Ray Model in 3D} -- When a wave interacts with an object of irregular shape in 3D, both WFC and the principal directions of the wavefronts evolve with each interaction, depending on the curvature properties of the object surface. Consequently, the primary challenge in extending VCRM to 3D case (VCRM-3D) lies in establishing the relationship between WFC of the incident, reflected, and refracted waves and the curvature of the object surface in 3D. Since the curvature properties of a surface can be fully described by a curvature matrix, the evolution of the wavefront can be expressed through its curvature matrices. By using differential geometry and phase matching, we derive (see Appendix for details) the relationship, termed the \emph{wavefront equation}, between the curvature matrices of the wavefronts before and after interaction $\Curv{Q}_i$ and $\Curv{Q}'$ (i.e., $\Curv{Q}_l$ or $\Curv{Q}_r$) as follows 
\vspace{-5mm}
\begin{figure}[!ht]
	\begin{center}
		\includegraphics[width=0.3\textwidth]{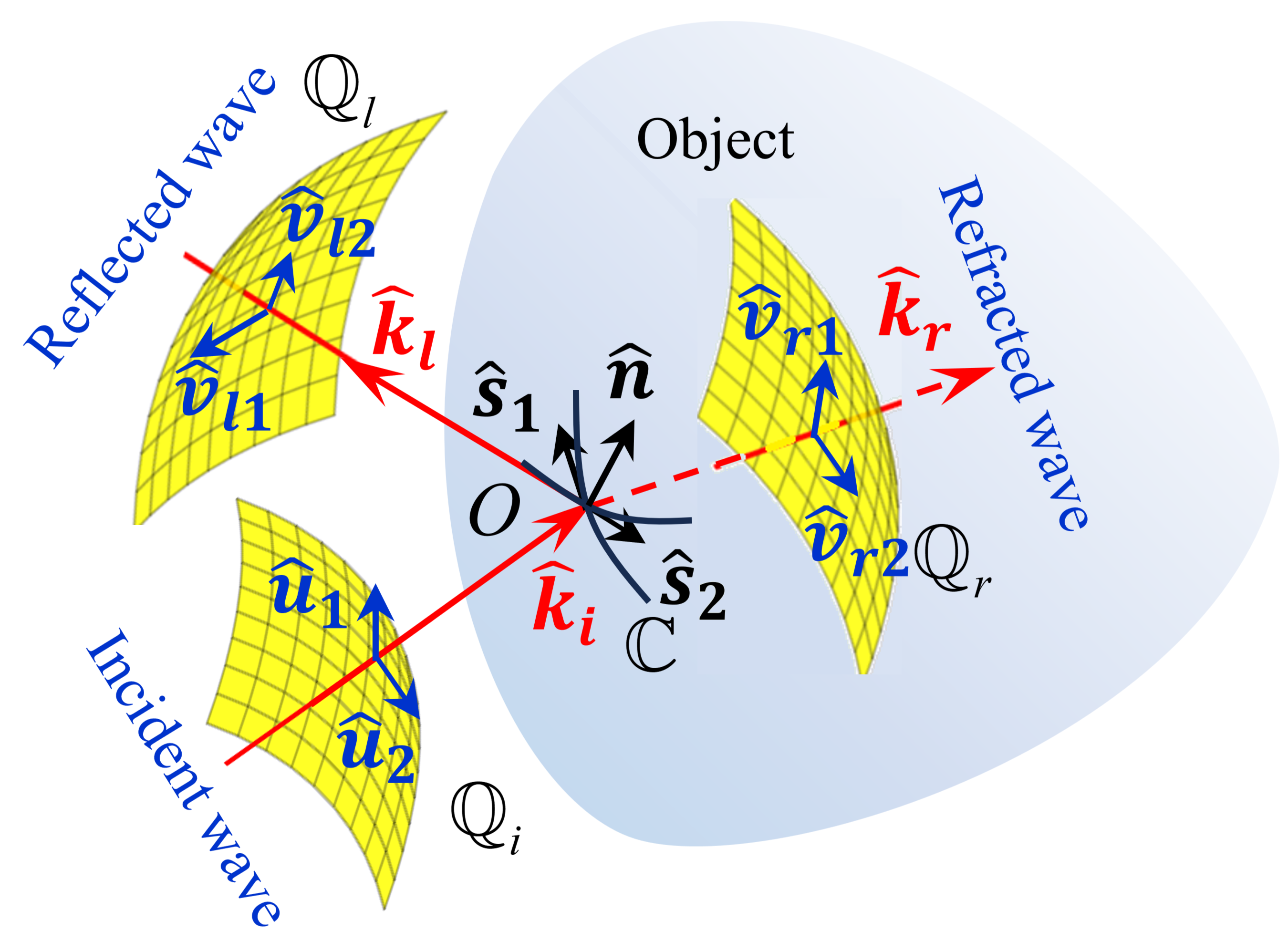}\vspace{-3mm}
		\caption{Schematic of the wavefront equation: An arbitrary wavefront interacts with a curved object surface. The curvature matrices $\Curv{C}$, $\Curv{Q}_i$, $\Curv{Q}_l$, and $\Curv{Q}_r$ are expressed in the respective bases: $(\umbs_1,\umbs_2)$, $(\umbu_1,\umbu_2)$, $(\umbv_{l1},\umbv_{l2})$, and $(\umbv_{r1},\umbv_{r2})$.}
		\label{fig:WaveFrontEqu3D}
	\end{center}
	\vspace{-8mm}
\end{figure}
\begin{equation}\label{eq:WavefrontEqVec}
	k'\Curv{P}' \Curv{Q}'P^{\prime T}=k_i\Curv{P}_i \Curv{Q}_i\Curv{P}_i^T+\Curv{C}\umbn\cdot(\mbk'-\mbk_i)
\end{equation}
where $\mbk$ is the wave vector and $k$ the wave number. The index $_i$ and the prime $'$ indicate quantities before (incident) and after the interaction (reflected or refracted wave), respectively. $\Curv{Q}$ denotes the wavefront curvature matrix and $\Curv{C}$ the object surface curvature matrix, and $\Curv{P}$ the projection matrix from the object surface basis $(\umbs_1,\umbs_2)$ to a wavefront basis.

Furthermore, to simplify the ray tracing in 3D we adopt Snell's law in vector form, which states that the tangent components of incident, reflected and refracted wave vectors are equal
 \begin{equation}
	k_{i\tau}=k_{l\tau}=k_{r\tau}
	\label{eq:SnellVec}
\end{equation}
This relationship naturally arises in the derivation of the wavefront equation (see Appendix). Although mathematically equivalent to the conventional Snell’s law, it offers greater convenience for 3D ray tracing.

Eqs. (\ref{eq:WavefrontEqVec}) and (\ref{eq:SnellVec}) are all we need to determine the directions and WFC of any rays. With some additional considerations described below, we can calculate the complex amplitude of all rays at any point.

\emph{Amplitude of a ray}, in the propagation of a wave and its interaction with an object, is affected by two factors: the Fresnel factor $\varepsilon_{X,p}$ and the divergence factor $\mathcal{D}_p$, where $X$ denotes the polarization state ($\perp$ or $\parallel$) and $p$ the ray order. The wave amplitude at a point $M$ along a ray is then given by
\begin{equation}\label{eq:IntVCRM}
	A=A_0|\varepsilon_{X,p}|\sqrt{\mathcal{D}_p}
\end{equation}
where $A_0$ is the amplitude at the first incident point with $\mathcal{D}_p$ given by
\begin{equation} \label{eq:DivFactVCRM}
	\mathcal{D}_p=\left|\frac{\kappa_{G2}}{\kappa_{G1}'}\cdot\frac{\kappa_{G3}}{\kappa_{G2}'} \cdots\frac{\kappa_{GM}}{\kappa_{G(p+1)}'}\right|
\end{equation}
$\kappa_{Gq}$ and $\kappa_{Gq}'$ denote the Gaussian curvatures of the incident and refracted/reflected wavefronts at the $q^{th}$ interaction point, and $\kappa_{GM}$ the Gaussian curvature of the wavefront at observation point $M$. 
The factor $\varepsilon_{X,p}$ is calculated by 
\begin{equation}\label{eq:epsilon_VCRM}
	\varepsilon_{X,p}=\left\{
	\begin{array}{ll}
		r_{X,0} & p=0  \\
		t_{X,0}t_{X,p}\dspl{\prod_{q=1}^{p-1} r_{X,q}}& p\geq 1
	\end{array}
	\right.
\end{equation}
where $r_{X,q}$ and $t_{X,q}$ are the Fresnel reflection and refraction coefficients at the $q^{th}$ interaction, respectively.
In VCRM, both the reflection and refraction coefficients $r_X$ and $t_X$ are expressed in terms of normal components of the incident and refracted wave vectors $k_{in}$ and $k_{rn}$ as
\begin{equation}
	\label{eq:Fres}
	\begin{array}{cc}
		\displaystyle{r_\perp=\frac{k_{in}-k_{rn}}{k_{in}+k_{rn}},} \quad &
		\displaystyle{r_\parallel=\frac{m^2k_{in}-k_{rn}}{m^2k_{in}+k_{rn}}} \\[4mm]
		\displaystyle{t_\perp=\frac{2k_{in}}{k_{in}+k_{rn}},}  \quad &
		\displaystyle{t_\parallel=\frac{2mk_{in}}{m^2k_{in}+k_{rn}}}
	\end{array}
\end{equation}
In the general 3D case, the polarization state and the principal directions of the wavefront change at each interaction. In particular, the two components $\perp$ and $\parallel$ of $\varepsilon_{X,q}$ in Eq. (\ref{eq:epsilon_VCRM}) must be reevaluated at each interaction according to the incident plane. Therefore, $\mathcal{D}_p$ and $\varepsilon_{X,p}$ must be calculated step by step. In VCRM-Sym, since there is no cross-polarization, the two components of $\varepsilon_{X,p}$ can be calculated independently.

\emph{Phase of a ray} may change due to the optical path $\Phi_P$, reflection $\Phi_F$, and focal lines $\Phi_f$\cite{vandeHulst57,Berry2023}. $\Phi_P$ of a ray after $p+1$ interactions with the object is given by 
\begin{equation}\label{eq:PhasePath}
	\Phi_P=-\mbk_1\cdot\mbr_1+\mbk'_q\cdot\mbr_M
	+\sum_{q=2}^{p+1}\mbk_{q}\cdot(\mbr_{q}-\mbr_{q-1})
\end{equation}
where $\mbr_q$ and $\mbk_q$ are the position vector and wave vector of the incident wave at the $q^{th}$ interaction point, respectively, while $\mbk_q'$ is the wave vector after $q^{th}$ interaction and $\mbr_M$ the position vector of point $M$. $\Phi_F$ is the total phase shifts induced by the Fresnel coefficients. Each time a wave passes through a focal line (Fig. \ref{fig:RayTracing}), its phase experiences a jump of $\pi/2$ \cite{vandeHulst57,RenVCRMOptLett2011}, with the accumulation of these jumps yielding $\Phi_f$. This effect corresponds to the sign changes in WFC. In VCRM-3D, since both the polarization state and the principal directions of wavefront vary at each interaction, $\Phi_F$ and $\Phi_f$ must be calculated step by step. In contrast, in VCRM-Sym, where no cross polarization occurs, $\Phi_F$ simplifies to the argument of Fresnel factor, i.e. $\Phi_F=\arg(\varepsilon_{X,p})$. The calculation of $\Phi_f$ reduces to counting the number $N_f$ of WFC sign changes, leading to $\Phi_f=N_f\frac{\pi}{2}$.

If the incident wave is not a plane wave, its phase $\Phi_i$ at the first incident point must be taken into account. Therefore, the total phase of a ray at $M$ is given by
\begin{equation}\label{eq:PhaseTotal}
  \Phi(M)=\Phi_i+\Phi_F+\Phi_P+\Phi_f
\end{equation}

Eqs. (\ref{eq:IntVCRM}) and (\ref{eq:PhaseTotal}) allow the calculation of the amplitudes and phases of all rays at any point using VCRM. 
\vspace{-8mm}
\begin{figure}[!ht]
	\begin{center}
		\includegraphics[width=0.4\textwidth]{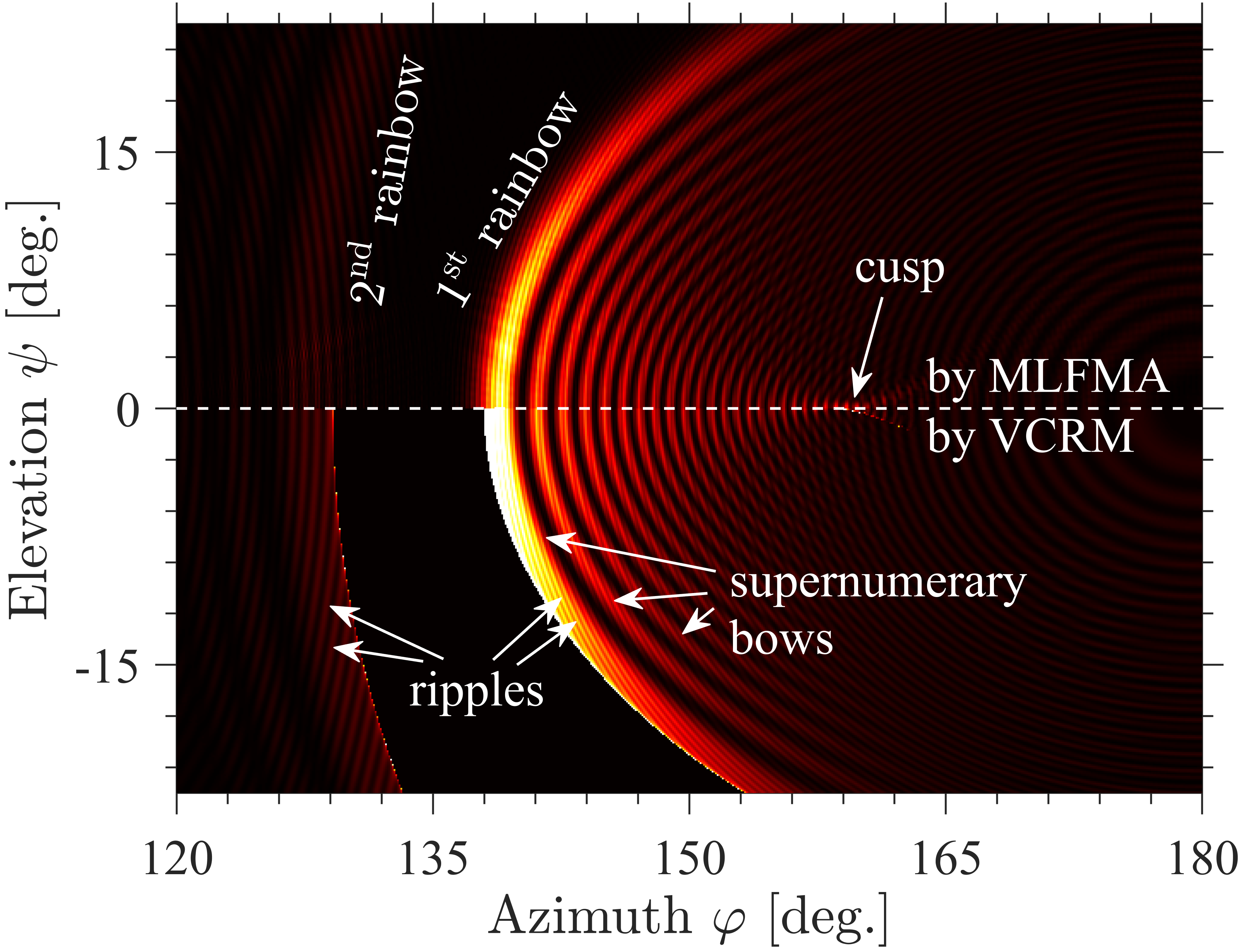}
		\caption{Comparison of scattering patterns computed using VCRM (lower part) and MLFMA (upper part) for an oblate spheroid of water ($m=1.333$) with semi-axes $a=b=100~\upmu$m, $c=90~\upmu$m. The incident plane wave, with wavelength $\lambda=0.6328~\upmu$m, propagates along $x$-axis.}
		\label{fig:3D-VCRM-MLFMA}
	\end{center}
\end{figure}

\begin{figure*}[!ht]
	\begin{center}
		\includegraphics[width=0.9\textwidth]{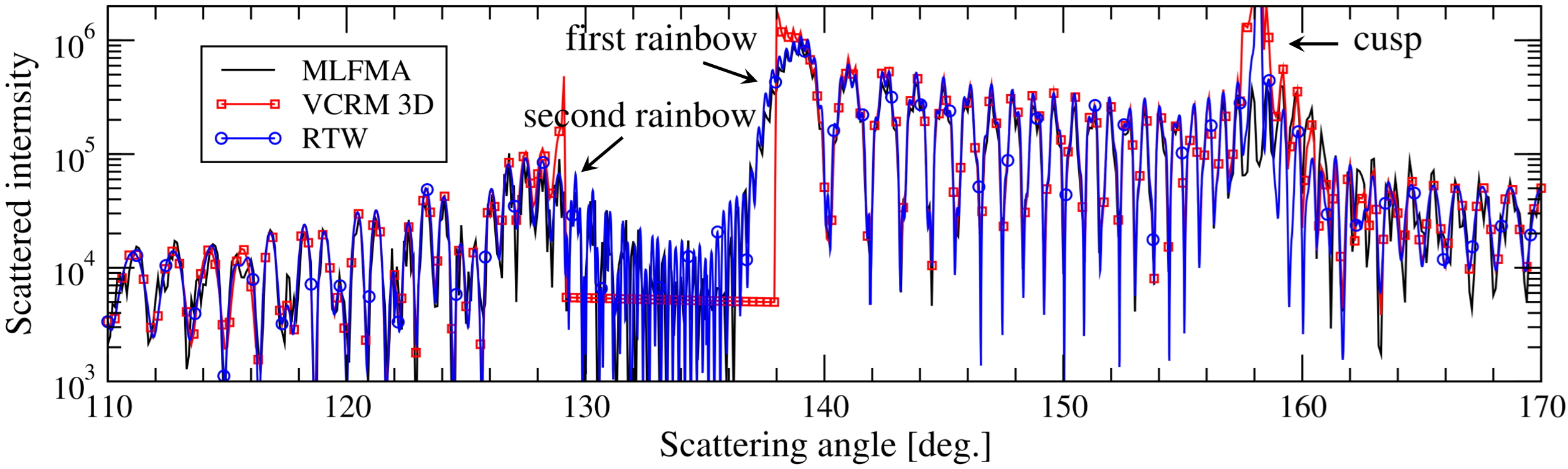}
		\caption{Numerical comparison of the scattering diagrams in the plane $z=0$ calculated by three methods MLFMA, VCRM and RTW with the same parameters as in Fig. \ref{fig:3D-VCRM-MLFMA}.}
		\label{fig:2D-VCRM-MLFMA-RTW}
	\end{center}
\end{figure*}

The scattering patterns of an oblate ellipsoid have been successfully simulated using VCRM-3D \cite{Duan2021OL}. However, two notable discrepancies arise when compared to experimental results (Fig. 6 in \cite{Duan2021OL}). The first, observed near the rainbow angles and cusp, is attributed to caustics and can be corrected using physical optics (PO), which will be discussed later. The second, at large angles, is likely due to experimental uncertainty. To clarify this issue, we compare the results of VCRM with those of MLFMA in Fig. \ref{fig:3D-VCRM-MLFMA} and Fig. \ref{fig:2D-VCRM-MLFMA-RTW}. It is evident that the agreement is very good everywhere including the fine structure, except in the vicinity of caustics (rainbow angles and cusp). The discrepancy at large angles is confirmed to be due to the experimental uncertainty. MLFMA computation (Fig. \ref{fig:3D-VCRM-MLFMA}) with an angular resolution of $0.1^\circ$ required 997 GB of memory (Intel Xeon Platinum 8276 CPU) and 40 hours of CPU time using 160 threads. In contrast, VCRM computation on a PC (Intel i9-13900ks, 32 GB RAM) took only 9.4 minutes for an angular resolution of $0.01^\circ$.

\textbf{Diffraction effect correction based on VCRM} -- VCRM significantly enhances the flexibility and accuracy of ray-based models. However, it still fails to predict correctly the field near singularity points (e.g., caustics). Since VCRM computes rigorously the complex amplitude \emph{at any point} directly for a given object within the ray framework, these singularities can be corrected by incorporating wave effects in the framework of RTW. This requires a specific strategy, which will be described later. To illustrate the methodology and performance of RTW, we present two typical examples.

\begin{table}[!ht]
	\caption{Peak positions (in deg.) of supernumerary bows for a spherical particle of radius $a$ and refractive index $m=1.333$ illuminated by a plane wave of wavelength $\lambda=0.6328~\upmu$m.}
	\label{tab:Rainbowpeaks}
	\begin{tabular}{r|r|l|r||r|r|l|r}
		\hline\hline
		\multicolumn{4}{c||}{$a=50~\upmu$m} &	\multicolumn{4}{c}{$a=500~\upmu$m}\\\hline
		$K$ & Airy   & RTW & Debye	&$K$	&Airy	&RTW	&Debye\\ \hline\hline
		0 & 139.50	 & \emph{139.47}	 & 139.47	 & 0	 & 138.26	 & \emph{138.26}	 & 138.26\\ \hline
		1 &	142.98	 & 142.88	 & 142.87	 & 1	 & 139.01	 & 139.00	 & 139.00\\ \hline
		4 & 149.37	 & 149.01	 & 148.99	 & 41	 & 149.14	 & 148.81	 & 148.81\\ \hline
		8 & 155.73	 & 154.91	 & 154.88	 & 81	 & 155.55	 & 154.76	 & 154.76\\ \hline
		12 & 161.10 & 159.77	 & 159.72	 & 121	 & 160.95	 & 159.64	 & 159.64\\ \hline\hline
	\end{tabular}
\end{table}

The rainbow is a remarkable wonder of nature, and its cause has always been curiously sought after by great minds\cite{Boyer1987,KharePRL1974,Adam2002Rainbow}. Among these, Airy theory stands as one of the most renowned and has served for various scientific studies. However, its precision and applicability have been questioned for long time \cite{Boitel1888,LeeAO98}. Thanks to VCRM-Sym we have demon\-strated\cite{Zhang2022} that its combination with PO accurately predicts the rainbow diagram of a spherical drop. This work provides clear answers to several queries on Airy theory since 19$^{th}$ centuries\cite{Boitel1888,Boyer1987,Adam2002Rainbow}, including: (1). its failure for parallel polarization\cite{KharePRL1974,Konnen:79} due to the omission of the phase jump at Brewster angle; (2). its discrepancy with rigorous theory\cite{Hovenac92} arising  from the approximation in the calculation of amplitude and phase; and (3) the question of its applicable particle size\cite{vandeHulst57,LeeAO98} is fundamentally misguided. To clarify the last point, we compare in
Tab. \ref{tab:Rainbowpeaks} the supernumerary peak positions calculated by Airy theory and RTW with rigorous Debye theory\cite{hovenac93}. VCRM-Sym predicts accurately all the supernumerary peaks ($K\ge1$) except the main one ($K=0$) which requires the combination of VCRM-Sym and PO. It reveals also that, contrary to common belief\cite{vandeHulst57}, the discrepancy of Airy theory is nearly independent of the particle radius. 

In the framework of RTW, the approach described above can be readily extended to non-spherical objects.  Fig. \ref{fig:2D-VCRM-MLFMA-RTW} presents a quantitative comparison of the scattering diagram in the symmetric plane ($\psi=0$ in Fig.\ref{fig:3D-VCRM-MLFMA}). The result of RTW has been calculated by simple integration. RTW closely matches MLFMA, both near the rainbow angles and in the Alexander region. The discrepancy around $160^\circ$ arises from 3D caustics requiring 2D integration. This issue is currently under study using singularity theory\cite{Berry80,Berry2023} and uniform approximation\cite{Berry1969}.

\textbf{Strategy of Ray Theory of Waves} -- 
VCRM enables the rigorous calculation of amplitudes and phases for rays of any order. In principle, the singularities inherent in ray models can be corrected by incorporating diffraction effects. However, an unconventional strategy is required to achieve this and obtain the total field at a given point or in a specific direction.

In conventional practice, a physical problem is considered solved if the relation between a physical quantity $Y$ and its variables $\mb{x}=\{x_i\} (i=1,2,\cdots)$ is established rigorously or approximately in the form $Y=\mathcal{F}(\mb{x};\mb{\alpha})$, where $\mb{\alpha}=\{\alpha_j\} (j=1,2,\cdots)$ are the problem related parameters and the operator $\mathcal{F}$ can be in usual or special functions, a differential or integral operator, or their combination. The Mie theory, Airy theory and aberration theory are typical examples.
Otherwise, An implicit relation between $Y$ and $\mb{x}$ is established in the form $\mathcal{F}(Y,\mb{x};\mb{\alpha})=0$ then the equation is solved numerically. This is largely used in EM computation.

RTW follows neither of the above strategies. \!Instead, we express $Y$ and $\mb{x}$ through intermediate variables $\mb{t}={t_i}$, such that $Y=\mathcal{F}_1(\mb{t};\mb{\alpha})$ and $\mb{x}=\mathcal{F}_2(\mb{t};\mb{\alpha})$, then establish their relationship numerically. For instance, to obtain the scattering field, we compute the amplitudes $A$ and phase $\Phi$, i.e., $Y=(A,\Phi)$, and the emergent ray directions $\mb{x}=(\theta_e,\phi_e)$ as functions of the incident rays $\mb{t}=(\theta_i,\phi_i)$ and object properties $\mb{\alpha}$. Interpolation is then applied to compute the total complex amplitude of all rays in the region of interest, with \cite{Zhang2022} or without \cite{Duan2023OE} wave effect correction. Contrary to conventional approaches, which often rely on series expansions (e.g., Airy theory and aberration theory) to approximate physical quantities—leading to tedious mathematics—RTW directly computes these quantities rigorously within the ray framework. This approach is both computationally efficient and highly flexible, preserving the option for series representation when needed.

\textbf{Conclusions} --
The fundamentals of the \emph{Ray Theory of Waves} (RTW) are established, and its specific strategy is described. Its flexibility, accuracy, and efficiency are demonstrated through comparisons with MLFMA. A typical application to the rainbow scattering of a spherical drop clarifies several aspects of Airy theory, while its direct extension to non-spherical drops is validated through comparison with numerical methods. RTW offers broad application potentials, including optical metrology in fluid mechanics, electromagnetic (EM) computation, freeform optics, and computer graphics. The correction of singularities in 3D scattering is currently under development.

\textbf{Acknowledgments}
We thank Prof. B. Pouligny, Prof. L. Méès, Pro. J. C. Loudet, Dr. S. Idlahcen, and Dr. X. Qiao for fruitful discussion. This work was partially supported by the National Natural Science Foundation of China under Grants 62231003 and 62205262, the Agence Nationale de la Recherche (ANR-13-BS090008-01 AMO-COPS) of France.

\appendix
\section{Appendix: Derivation of the Wavefront Equation}

\begin{figure}[!ht]
	\begin{center}
		\includegraphics[width=0.4\textwidth]{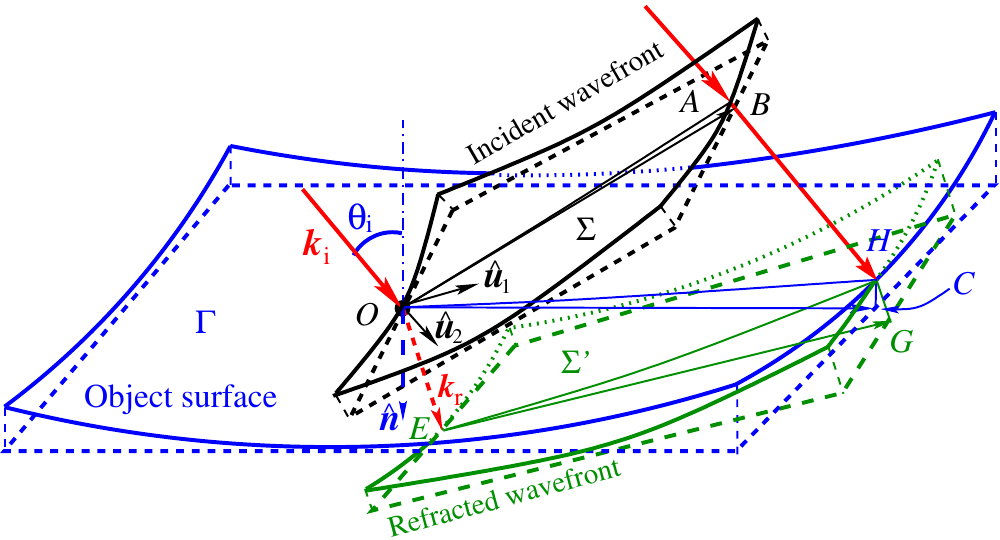}
		\caption{Schema of wavefronts and object surface: a wave of arbitrary shape incident on a curved surface. The dashed lines represent the tangent plane of corresponding surface.}
		\label{fig:FrontCurvCurv3D}
	\end{center}
	\vspace{-8mm}
\end{figure}

Consider a ray incident on an object surface $\Gamma$ at a point $O$ (Fig. \ref{fig:FrontCurvCurv3D}). When an adjacent ray passing by the point $A$ on the incident wavefront $\Sigma$ reaches the surface at $H$, the ray refracted at $O$ arrives at $E$ on the refracted wavefront $\Sigma'$. We note $B$ as the projection point of $A$ on the tangent plane of $\Sigma$, $C$ and $G$ the projection points of $H$ respectively on the tangent planes of $\Gamma$ and $\Sigma'$. The curvature of the surface $\Sigma$ is described by its curvature matrix $\Curv{Q}_i$ in the base $(\umbu_1,\umbu_2)$. Similarly, the curvatures of $\Sigma'$ and $\Gamma$ are described respectively by their curvature matrices $\Curv{Q}_r$ and $\Curv{C}$ in the corresponding bases $(\umbs_1,\umbs_2)$ and $(\umbv_1,\umbv_2)$
(not presented in the figure for clarity). According to the differential geometry, in the vicinity of tangent point, the distance between a point on the curved surface and the projected point on the tangent plane can be expressed as function of the curvature matrix. So, the infinitesimal distances $\delta_i=\overline{AB}$, $\delta_r=\overline{HG}$ and $\delta_c=\overline{HC}$ are given by
\begin{equation*}
	\delta_i=\demi \mbt_i\Curv{Q}_i \mbt_i^T, \quad
	\delta_r=\demi \mbt_r\Curv{Q}_r \mbt_r^T , ~~
	\delta_c=\demi \mbt_c\Curv{C} \mbt_c^T
\end{equation*}
where $\mbt_i=\overrightarrow{OB}$, $\mbt_r=\overrightarrow{EG}$ and  $\mbt_c=\overrightarrow{OC}$. The prime $^T$ stands for the transpose of a matrix. Note that the quantities $\delta$ and $\mbt$ are all infinitely small, but the differential symbol $d$ of $d\delta$ and $d\mbt$ is omitted here for clarity.

The phase difference between the wavefronts $\Sigma$ and $\Sigma'$ being constant, so $k_i\overline{AH}=k_r\overline{OE}$. On the other hand, from the geometry we have
$\overline{AH}=\delta_i+\mbt_c\cdot\umbk_i-\delta_c\umbn\cdot\umbk_i $ and
$\overline{OE}=\delta_r+\mbt_c\cdot\umbk_r-\delta_c\umbn\cdot\umbk_r$. These lead to
\begin{equation*}
	k_i(\delta_i+\mbt_c\cdot\umbk_i-\delta_c\umbn\cdot\umbk_i)=
	k_r(\delta_r+\mbt_c\cdot\umbk_r-\delta_c\umbn\cdot\umbk_r)
\end{equation*}

The terms $k_i\mbt_c\cdot\umbk_i$ and $k_r\mbt_c\cdot\umbk_r$ are independent of any curvatures, so they are equal, i.e. $k_{i,\tau}=k_{r,\tau}$. This is also true for the reflected ray $k_{i,\tau}=k_{l,\tau}$. We establish thus the Snell law in vector form -- the tangent components of incident, reflected and refracted wave vectors are equal
$k_{i,\tau}=k_{l,\tau}=k_{r,\tau}$
The rest of the equation gives rise to
\begin{equation}\label{eq:WaveFrontEqn0}
	k_i\mbt_i\Curv{Q}_i\mbt_i^T-k_r\mbt_r\Curv{Q}_r\mbt_r^T
	=\umbn\cdot(\mbk_i-\mbk_r)\mbt_c\Curv{C}\mbt_c^T
\end{equation}

We express now $\mbt_i$ and $\mbt_r$ as function of $\mbt_c$ by using the projection relations
\begin{equation} \label{eq:ti_approx}
	\mbt_i=\Curv{P}_i \mbt_c+\mathcal{O}(\mbt_i^2), \quad \mbt_r=\Curv{P}_r\mbt_c+\mathcal{O}(\mbt_r^2)
\end{equation}
where $\Curv{P}_i$ and $\Curv{P}_r$ are the projection matrices of the surface base $(\umbs_1^i,\umbs_2^i)$ respectively on the incident and refracted wavefront bases $(\umbu_1,\umbu_2)$ and $(\umbv_1,\umbv_2)$
\begin{equation}\label{eq:Theta-r}
	\Curv{P}_i=\!\!\left(\!\!
	\begin{array}{cc}
		\umb{s}_1^i\cdot\umbu_1 & \umbs_1^i\cdot\umbu_2 \\
		\umb{s}_2^i\cdot\umbu_1 & \umbs_2^i\cdot\umbu_2 \\
	\end{array}
	\!\!\right),
	\quad
	\Curv{P}_r=\!\!\left(\!\!
	\begin{array}{cc}
		\umbs_1^i\cdot\umbv_1 & \umbs_1^i\cdot\umbv_2 \\
		\umbs_2^i\cdot\umbv_1 & \umbs_2^i\cdot\umbv_2 \\
	\end{array}
	\!\!\right)
\end{equation}
Neglecting the high order approximation term $\mathcal{O}(.)$, Eq. (\ref{eq:WaveFrontEqn0}) is written as
\begin{equation}\label{eq:WaveFrontEqn1}
	\mbt_c\left(k_i\Curv{P}_i\Curv{Q}_i\Curv{P}_i^T-k_r\Curv{P}_r\Curv{Q}_r\Curv{P}_r^T\right)\mbt_c^T
	=\umbn\cdot(\mbk_i-\mbk_r)\mbt_c\Curv{C} \mbt_c^T
\end{equation}
$\mbt_c$ being arbitrary, we deduce finally the \emph{wavefront equation} for the refracted ray as
\begin{equation}\label{eq:WavefrontEqVecRefra}
	k_r\Curv{P}_r \Curv{Q}_r\Theta_r^T=k_i\Curv{P}_i \Curv{Q}_i\Curv{P}_i^T+\Curv{C}\umbn\cdot(\mbk_r-\mbk_i)
\end{equation}
The same procedure applies to the reflected ray. The general wavefront equation, relating wavefronts before and after interaction, is given in Eq. (\ref{eq:WavefrontEqVec}). The equations in \cite{Deschamps72} and \cite{JamesGTD80} correspond to the special case where a basis vector lies in the incident plane. For symmetric problems, Synge's relations naturally follow from Eq. (\ref{eq:WavefrontEqVec}).

\bibliography{../../../../BIB/LightScattering}

\begin{thebibliography}{44}%
\makeatletter
\providecommand \@ifxundefined [1]{%
 \@ifx{#1\undefined}
}%
\providecommand \@ifnum [1]{%
 \ifnum #1\expandafter \@firstoftwo
 \else \expandafter \@secondoftwo
 \fi
}%
\providecommand \@ifx [1]{%
 \ifx #1\expandafter \@firstoftwo
 \else \expandafter \@secondoftwo
 \fi
}%
\providecommand \natexlab [1]{#1}%
\providecommand \enquote  [1]{``#1''}%
\providecommand \bibnamefont  [1]{#1}%
\providecommand \bibfnamefont [1]{#1}%
\providecommand \citenamefont [1]{#1}%
\providecommand \href@noop [0]{\@secondoftwo}%
\providecommand \href [0]{\begingroup \@sanitize@url \@href}%
\providecommand \@href[1]{\@@startlink{#1}\@@href}%
\providecommand \@@href[1]{\endgroup#1\@@endlink}%
\providecommand \@sanitize@url [0]{\catcode `\\12\catcode `\$12\catcode
  `\&12\catcode `\#12\catcode `\^12\catcode `\_12\catcode `\%12\relax}%
\providecommand \@@startlink[1]{}%
\providecommand \@@endlink[0]{}%
\providecommand \url  [0]{\begingroup\@sanitize@url \@url }%
\providecommand \@url [1]{\endgroup\@href {#1}{\urlprefix }}%
\providecommand \urlprefix  [0]{URL }%
\providecommand \Eprint [0]{\href }%
\providecommand \doibase [0]{https://doi.org/}%
\providecommand \selectlanguage [0]{\@gobble}%
\providecommand \bibinfo  [0]{\@secondoftwo}%
\providecommand \bibfield  [0]{\@secondoftwo}%
\providecommand \translation [1]{[#1]}%
\providecommand \BibitemOpen [0]{}%
\providecommand \bibitemStop [0]{}%
\providecommand \bibitemNoStop [0]{.\EOS\space}%
\providecommand \EOS [0]{\spacefactor3000\relax}%
\providecommand \BibitemShut  [1]{\csname bibitem#1\endcsname}%
\let\auto@bib@innerbib\@empty
\bibitem [{\citenamefont {van~de Hulst}(1957)}]{vandeHulst57}%
  \BibitemOpen
  \bibfield  {author} {\bibinfo {author} {\bibfnamefont {H.~C.}\ \bibnamefont
  {van~de Hulst}},\ }\href@noop {} {\emph {\bibinfo {title} {Light scattering
  by small particles}}}\ (\bibinfo  {publisher} {John Wiley \& Sons},\ \bibinfo
  {address} {New York},\ \bibinfo {year} {1957})\BibitemShut {NoStop}%
\bibitem [{\citenamefont {Khare}\ and\ \citenamefont
  {Nussenzveig}(1974)}]{KharePRL1974}%
  \BibitemOpen
  \bibfield  {author} {\bibinfo {author} {\bibfnamefont {V.}~\bibnamefont
  {Khare}}\ and\ \bibinfo {author} {\bibfnamefont {H.~M.}\ \bibnamefont
  {Nussenzveig}},\ }\bibfield  {title} {\bibinfo {title} {Theory of the
  rainbow},\ }\href@noop {} {\bibfield  {journal} {\bibinfo  {journal} {Phys.
  Rev. Lett.}\ }\textbf {\bibinfo {volume} {33}},\ \bibinfo {pages} {976}
  (\bibinfo {year} {1974})}\BibitemShut {NoStop}%
\bibitem [{\citenamefont {Nye}(1984)}]{NyeNature1984}%
  \BibitemOpen
  \bibfield  {author} {\bibinfo {author} {\bibfnamefont {J.~F.}\ \bibnamefont
  {Nye}},\ }\bibfield  {title} {\bibinfo {title} {Rainbow scattering from
  spheroidal drops - an explanation of the hyperbolic umbilic foci},\
  }\href@noop {} {\bibfield  {journal} {\bibinfo  {journal} {Nature}\ }\textbf
  {\bibinfo {volume} {312(5994)}},\ \bibinfo {pages} {531} (\bibinfo {year}
  {1984})}\BibitemShut {NoStop}%
\bibitem [{\citenamefont {Berry}\ and\ \citenamefont
  {Upstill}(1980)}]{Berry80}%
  \BibitemOpen
  \bibfield  {author} {\bibinfo {author} {\bibfnamefont {M.}~\bibnamefont
  {Berry}}\ and\ \bibinfo {author} {\bibfnamefont {C.}~\bibnamefont
  {Upstill}},\ }\bibfield  {title} {\bibinfo {title} {Catastrophe optics:
  Morphologies of caustics and their diffraction patterns},\ }\href@noop {}
  {\bibfield  {journal} {\bibinfo  {journal} {Prog. Opt.}\ }\textbf {\bibinfo
  {volume} {18}},\ \bibinfo {pages} {257} (\bibinfo {year} {1980})}\BibitemShut
  {NoStop}%
\bibitem [{\citenamefont {Taflove}\ and\ \citenamefont
  {Hagness}(2015)}]{Taflove2015}%
  \BibitemOpen
  \bibfield  {author} {\bibinfo {author} {\bibfnamefont {A.}~\bibnamefont
  {Taflove}}\ and\ \bibinfo {author} {\bibfnamefont {S.~C.}\ \bibnamefont
  {Hagness}},\ }\href@noop {} {\emph {\bibinfo {title} {Computational
  Electrodynamics: The Finite-difference Time-domain Method}}}\ (\bibinfo
  {publisher} {Artech House},\ \bibinfo {year} {2015})\BibitemShut {NoStop}%
\bibitem [{\citenamefont {Yang}\ \emph {et~al.}(2019)\citenamefont {Yang},
  \citenamefont {Wu}, \citenamefont {Gao},\ and\ \citenamefont
  {Sheng}}]{Yang2019}%
  \BibitemOpen
  \bibfield  {author} {\bibinfo {author} {\bibfnamefont {M.-L.}\ \bibnamefont
  {Yang}}, \bibinfo {author} {\bibfnamefont {B.-Y.}\ \bibnamefont {Wu}},
  \bibinfo {author} {\bibfnamefont {H.-W.}\ \bibnamefont {Gao}},\ and\ \bibinfo
  {author} {\bibfnamefont {X.-Q.}\ \bibnamefont {Sheng}},\ }\bibfield  {title}
  {\bibinfo {title} {A ternary parallelization approach of {MLFMA} for solving
  electromagnetic scattering problems with over 10 billion unknowns},\ }\href
  {https://doi.org/10.1109/tap.2019.2927660} {\bibfield  {journal} {\bibinfo
  {journal} {IEEE Trans. Antennas Propag.}\ }\textbf {\bibinfo {volume} {67}},\
  \bibinfo {pages} {6965} (\bibinfo {year} {2019})}\BibitemShut {NoStop}%
\bibitem [{\citenamefont {Notaros}(2024)}]{NotarosIEEE-APM75years}%
  \BibitemOpen
  \bibfield  {author} {\bibinfo {author} {\bibfnamefont {B.~M.}\ \bibnamefont
  {Notaros}},\ }\bibfield  {title} {\bibinfo {title} {75 years of {IEEE AP-S}
  research in computational electromagnetics: A view on the discipline and its
  history, current state, and future prospects},\ }\href@noop {} {\bibfield
  {journal} {\bibinfo  {journal} {IEEE AP Mag}\ }\textbf {\bibinfo {volume}
  {66}},\ \bibinfo {pages} {68} (\bibinfo {year} {2024})}\BibitemShut {NoStop}%
\bibitem [{\citenamefont {Glassner}(1989)}]{GlassnerRayTr1989}%
  \BibitemOpen
  \bibfield  {author} {\bibinfo {author} {\bibfnamefont {A.}~\bibnamefont
  {Glassner}},\ }\href@noop {} {\emph {\bibinfo {title} {An Introduction To Ray
  Tracing}}}\ (\bibinfo  {publisher} {Academic Press},\ \bibinfo {year}
  {1989})\BibitemShut {NoStop}%
\bibitem [{\citenamefont {Pharr}\ \emph {et~al.}(2004)\citenamefont {Pharr},
  \citenamefont {Jakob},\ and\ \citenamefont {Humphreys}}]{PharrBook2004}%
  \BibitemOpen
  \bibfield  {author} {\bibinfo {author} {\bibfnamefont {M.}~\bibnamefont
  {Pharr}}, \bibinfo {author} {\bibfnamefont {W.}~\bibnamefont {Jakob}},\ and\
  \bibinfo {author} {\bibfnamefont {G.}~\bibnamefont {Humphreys}},\ }\href@noop
  {} {\emph {\bibinfo {title} {Physically Based Rendering}}}\ (\bibinfo
  {publisher} {MIT Press},\ \bibinfo {year} {2004})\BibitemShut {NoStop}%
\bibitem [{\citenamefont {Marrs}\ \emph {et~al.}(2021)\citenamefont {Marrs},
  \citenamefont {Shirley},\ and\ \citenamefont {Wald}}]{RayTrGemsII2021}%
  \BibitemOpen
  \bibfield  {author} {\bibinfo {author} {\bibfnamefont {A.}~\bibnamefont
  {Marrs}}, \bibinfo {author} {\bibfnamefont {P.}~\bibnamefont {Shirley}},\
  and\ \bibinfo {author} {\bibfnamefont {I.}~\bibnamefont {Wald}},\ }\href@noop
  {} {\emph {\bibinfo {title} {Ray Tracing Gems II}}}\ (\bibinfo  {publisher}
  {{NVIDIA} \& Apress},\ \bibinfo {year} {2021})\BibitemShut {NoStop}%
\bibitem [{\citenamefont {Holl}\ and\ \citenamefont
  {Reinhard†}(2017)}]{HollPRL2017}%
  \BibitemOpen
  \bibfield  {author} {\bibinfo {author} {\bibfnamefont {P.~M.}\ \bibnamefont
  {Holl}}\ and\ \bibinfo {author} {\bibfnamefont {F.}~\bibnamefont
  {Reinhard†}},\ }\bibfield  {title} {\bibinfo {title} {Holography of {Wi-fi}
  radiation},\ }\href@noop {} {\bibfield  {journal} {\bibinfo  {journal} {Phys.
  Rev. Lett.}\ }\textbf {\bibinfo {volume} {118}},\ \bibinfo {pages} {183901}
  (\bibinfo {year} {2017})}\BibitemShut {NoStop}%
\bibitem [{\citenamefont {Pan}\ \emph {et~al.}(2024)\citenamefont {Pan},
  \citenamefont {Trusler}, \citenamefont {Jin},\ and\ \citenamefont
  {Zhang}}]{PanNature2024}%
  \BibitemOpen
  \bibfield  {author} {\bibinfo {author} {\bibfnamefont {Z.}~\bibnamefont
  {Pan}}, \bibinfo {author} {\bibfnamefont {J.~P.~M.}\ \bibnamefont {Trusler}},
  \bibinfo {author} {\bibfnamefont {Z.}~\bibnamefont {Jin}},\ and\ \bibinfo
  {author} {\bibfnamefont {K.}~\bibnamefont {Zhang}},\ }\bibfield  {title}
  {\bibinfo {title} {Interfacial property determination from dynamic
  pendant-drop characterizations},\ }\href@noop {} {\bibfield  {journal}
  {\bibinfo  {journal} {Nature Protocols}\ } (\bibinfo {year}
  {2024})}\BibitemShut {NoStop}%
\bibitem [{\citenamefont {Rolland}\ \emph {et~al.}(2021)\citenamefont
  {Rolland}, \citenamefont {Davies}, \citenamefont {Suleski}, \citenamefont
  {Evans}, \citenamefont {Bauer}, \citenamefont {Lambropoulos},\ and\
  \citenamefont {Falaggis}}]{RollandOptica2021}%
  \BibitemOpen
  \bibfield  {author} {\bibinfo {author} {\bibfnamefont {J.~P.}\ \bibnamefont
  {Rolland}}, \bibinfo {author} {\bibfnamefont {M.~A.}\ \bibnamefont {Davies}},
  \bibinfo {author} {\bibfnamefont {T.~J.}\ \bibnamefont {Suleski}}, \bibinfo
  {author} {\bibfnamefont {C.}~\bibnamefont {Evans}}, \bibinfo {author}
  {\bibfnamefont {A.}~\bibnamefont {Bauer}}, \bibinfo {author} {\bibfnamefont
  {J.~C.}\ \bibnamefont {Lambropoulos}},\ and\ \bibinfo {author} {\bibfnamefont
  {K.}~\bibnamefont {Falaggis}},\ }\bibfield  {title} {\bibinfo {title}
  {Freeform optics for imaging},\ }\href@noop {} {\bibfield  {journal}
  {\bibinfo  {journal} {Optica}\ }\textbf {\bibinfo {volume} {8}},\ \bibinfo
  {pages} {161} (\bibinfo {year} {2021})}\BibitemShut {NoStop}%
\bibitem [{\citenamefont {Yang}\ \emph {et~al.}(2021)\citenamefont {Yang},
  \citenamefont {Shen}, \citenamefont {Ding}, \citenamefont {Tao},
  \citenamefont {Zheng}, \citenamefont {Wu}, \citenamefont {Li},\ and\
  \citenamefont {Wu}}]{YangOE2021}%
  \BibitemOpen
  \bibfield  {author} {\bibinfo {author} {\bibfnamefont {L.}~\bibnamefont
  {Yang}}, \bibinfo {author} {\bibfnamefont {F.}~\bibnamefont {Shen}}, \bibinfo
  {author} {\bibfnamefont {Z.}~\bibnamefont {Ding}}, \bibinfo {author}
  {\bibfnamefont {X.}~\bibnamefont {Tao}}, \bibinfo {author} {\bibfnamefont
  {Z.}~\bibnamefont {Zheng}}, \bibinfo {author} {\bibfnamefont
  {F.}~\bibnamefont {Wu}}, \bibinfo {author} {\bibfnamefont {Y.}~\bibnamefont
  {Li}},\ and\ \bibinfo {author} {\bibfnamefont {R.}~\bibnamefont {Wu}},\
  }\bibfield  {title} {\bibinfo {title} {Freeform optical design of beam
  shaping systems with variable illumination properties},\ }\href@noop {}
  {\bibfield  {journal} {\bibinfo  {journal} {Opt. Express}\ }\textbf {\bibinfo
  {volume} {29}},\ \bibinfo {pages} {31993} (\bibinfo {year}
  {2021})}\BibitemShut {NoStop}%
\bibitem [{\citenamefont {Sheng}\ and\ \citenamefont
  {Song}(2012)}]{ShengSong2012}%
  \BibitemOpen
  \bibfield  {author} {\bibinfo {author} {\bibfnamefont {X.~Q.}\ \bibnamefont
  {Sheng}}\ and\ \bibinfo {author} {\bibfnamefont {W.}~\bibnamefont {Song}},\
  }\href@noop {} {\emph {\bibinfo {title} {Essentials of computational
  electromagnetics}}}\ (\bibinfo  {publisher} {Wiley},\ \bibinfo {address}
  {Singapore},\ \bibinfo {year} {2012})\BibitemShut {NoStop}%
\bibitem [{\citenamefont {Wu}\ and\ \citenamefont {Chew}(2016)}]{WuYM2016}%
  \BibitemOpen
  \bibfield  {author} {\bibinfo {author} {\bibfnamefont {Y.~M.}\ \bibnamefont
  {Wu}}\ and\ \bibinfo {author} {\bibfnamefont {W.~C.}\ \bibnamefont {Chew}},\
  }\bibfield  {title} {\bibinfo {title} {The modern high frequency methods for
  solving electromagnetic scattering problems},\ }\href@noop {} {\bibfield
  {journal} {\bibinfo  {journal} {PIER}\ }\textbf {\bibinfo {volume} {156}},\
  \bibinfo {pages} {63} (\bibinfo {year} {2016})}\BibitemShut {NoStop}%
\bibitem [{\citenamefont {Song}\ \emph {et~al.}(1997)\citenamefont {Song},
  \citenamefont {Lu},\ and\ \citenamefont {Chew}}]{SongIEEE1997}%
  \BibitemOpen
  \bibfield  {author} {\bibinfo {author} {\bibfnamefont {J.~M.}\ \bibnamefont
  {Song}}, \bibinfo {author} {\bibfnamefont {C.~C.}\ \bibnamefont {Lu}},\ and\
  \bibinfo {author} {\bibfnamefont {W.~C.}\ \bibnamefont {Chew}},\ }\bibfield
  {title} {\bibinfo {title} {Multilevel fast multipole algorithm for
  electromagnetic scattering by large complex objects},\ }\href@noop {}
  {\bibfield  {journal} {\bibinfo  {journal} {IEEE Trans. Antennas Propag.}\
  }\textbf {\bibinfo {volume} {45}},\ \bibinfo {pages} {1488} (\bibinfo {year}
  {1997})}\BibitemShut {NoStop}%
\bibitem [{\citenamefont {Mishchenko}\ \emph {et~al.}(2000)\citenamefont
  {Mishchenko}, \citenamefont {Hovenier},\ and\ \citenamefont
  {Travis}}]{Mishchenko00}%
  \BibitemOpen
  \bibfield  {author} {\bibinfo {author} {\bibfnamefont {M.~I.}\ \bibnamefont
  {Mishchenko}}, \bibinfo {author} {\bibfnamefont {J.~W.}\ \bibnamefont
  {Hovenier}},\ and\ \bibinfo {author} {\bibfnamefont {L.~D.}\ \bibnamefont
  {Travis}},\ }\href@noop {} {\emph {\bibinfo {title} {Light Scattering by
  Nonspherical Particles: Theory, Measurements, and Applications}}}\ (\bibinfo
  {publisher} {Academic Press},\ \bibinfo {address} {san Diego},\ \bibinfo
  {year} {2000})\BibitemShut {NoStop}%
\bibitem [{\citenamefont {Yurkin}\ and\ \citenamefont
  {Hoekstra}(2007)}]{YurkinJQSRT2007DDA}%
  \BibitemOpen
  \bibfield  {author} {\bibinfo {author} {\bibfnamefont {M.}~\bibnamefont
  {Yurkin}}\ and\ \bibinfo {author} {\bibfnamefont {A.}~\bibnamefont
  {Hoekstra}},\ }\bibfield  {title} {\bibinfo {title} {The discrete dipole
  approximation: An overview and recent developments},\ }\href@noop {}
  {\bibfield  {journal} {\bibinfo  {journal} {J. Quantit. Spectros. Radiat.
  Transfer}\ }\textbf {\bibinfo {volume} {106}},\ \bibinfo {pages} {558}
  (\bibinfo {year} {2007})}\BibitemShut {NoStop}%
\bibitem [{\citenamefont {Marston}(1994)}]{Marston94}%
  \BibitemOpen
  \bibinfo {editor} {\bibfnamefont {P.~L.}\ \bibnamefont {Marston}},\ ed.,\
  \href@noop {} {\emph {\bibinfo {title} {Selected papers on geometrical
  aspects of scattering}}},\ Vol.\ \bibinfo {volume} {MS 89}\ (\bibinfo
  {publisher} {SPIE Milestone Series},\ \bibinfo {address} {United States},\
  \bibinfo {year} {1994})\BibitemShut {NoStop}%
\bibitem [{\citenamefont {Bi}\ and\ \citenamefont {Yang}(2013)}]{BiYang2013}%
  \BibitemOpen
  \bibfield  {author} {\bibinfo {author} {\bibfnamefont {L.}~\bibnamefont
  {Bi}}\ and\ \bibinfo {author} {\bibfnamefont {P.}~\bibnamefont {Yang}},\
  }\bibfield  {title} {\bibinfo {title} {Physical-geometric optics hybrid
  methods for computing the scattering and absorption properties of ice
  crystals and dust aerosols},\ }in\ \href@noop {} {\emph {\bibinfo {booktitle}
  {Light Scattering Reviews}}}\ (\bibinfo  {publisher} {Springer-Praxis},\
  \bibinfo {year} {2013})\ pp.\ \bibinfo {pages} {69--114}\BibitemShut
  {NoStop}%
\bibitem [{\citenamefont {Adam}(2017)}]{AdamBook2017}%
  \BibitemOpen
  \bibfield  {author} {\bibinfo {author} {\bibfnamefont {J.~A.}\ \bibnamefont
  {Adam}},\ }\href@noop {} {\emph {\bibinfo {title} {Rays, Waves, and
  Scattering}}}\ (\bibinfo  {publisher} {Princeton University Press},\ \bibinfo
  {year} {2017})\BibitemShut {NoStop}%
\bibitem [{\citenamefont {Dong}\ \emph {et~al.}(2022)\citenamefont {Dong},
  \citenamefont {Guo}, \citenamefont {Meng},\ and\ \citenamefont
  {Li}}]{Guo2022IEEE}%
  \BibitemOpen
  \bibfield  {author} {\bibinfo {author} {\bibfnamefont {C.}~\bibnamefont
  {Dong}}, \bibinfo {author} {\bibfnamefont {L.}~\bibnamefont {Guo}}, \bibinfo
  {author} {\bibfnamefont {X.}~\bibnamefont {Meng}},\ and\ \bibinfo {author}
  {\bibfnamefont {H.}~\bibnamefont {Li}},\ }\bibfield  {title} {\bibinfo
  {title} {An improved {GO-PO/PTD} hybrid method for {EM} scattering from
  electrically large complex targets},\ }\href@noop {} {\bibfield  {journal}
  {\bibinfo  {journal} {IEEE TAP}\ }\textbf {\bibinfo {volume} {70}},\ \bibinfo
  {pages} {12130} (\bibinfo {year} {2022})}\BibitemShut {NoStop}%
\bibitem [{\citenamefont {Berry}(2023)}]{Berry2023}%
  \BibitemOpen
  \bibfield  {author} {\bibinfo {author} {\bibfnamefont {M.~V.}\ \bibnamefont
  {Berry}},\ }\bibfield  {title} {\bibinfo {title} {The singularities of light:
  intensity, phase, polarisation},\ }\href@noop {} {\bibfield  {journal}
  {\bibinfo  {journal} {Light Sci Appl.}\ }\textbf {\bibinfo {volume} {12}},\
  \bibinfo {pages} {1} (\bibinfo {year} {2023})}\BibitemShut {NoStop}%
\bibitem [{\citenamefont {Berry}(1969)}]{Berry1969}%
  \BibitemOpen
  \bibfield  {author} {\bibinfo {author} {\bibfnamefont {M.~V.}\ \bibnamefont
  {Berry}},\ }\bibfield  {title} {\bibinfo {title} {Uniform approximation: a
  new concept in wave theory},\ }\href@noop {} {\bibfield  {journal} {\bibinfo
  {journal} {Sci. Progress}\ }\textbf {\bibinfo {volume} {57}},\ \bibinfo
  {pages} {43} (\bibinfo {year} {1969})}\BibitemShut {NoStop}%
\bibitem [{\citenamefont {Stamnes}(1986)}]{stamnes86}%
  \BibitemOpen
  \bibfield  {author} {\bibinfo {author} {\bibfnamefont {J.~J.}\ \bibnamefont
  {Stamnes}},\ }\href@noop {} {\emph {\bibinfo {title} {Wave in focal
  regions}}},\ \bibinfo {edition} {institute of physics publishing}\ ed.\
  (\bibinfo  {publisher} {Adam Hilger},\ \bibinfo {year} {1986})\BibitemShut
  {NoStop}%
\bibitem [{\citenamefont {K\"{o}nnen}\ and\ \citenamefont
  {de~Boer}(1979)}]{Konnen:79}%
  \BibitemOpen
  \bibfield  {author} {\bibinfo {author} {\bibfnamefont {G.~P.}\ \bibnamefont
  {K\"{o}nnen}}\ and\ \bibinfo {author} {\bibfnamefont {J.~H.}\ \bibnamefont
  {de~Boer}},\ }\bibfield  {title} {\bibinfo {title} {Polarized rainbow},\
  }\href@noop {} {\bibfield  {journal} {\bibinfo  {journal} {Appl. Opt.}\
  }\textbf {\bibinfo {volume} {18}},\ \bibinfo {pages} {1961} (\bibinfo {year}
  {1979})}\BibitemShut {NoStop}%
\bibitem [{\citenamefont {Synge}(1937)}]{Synge1937}%
  \BibitemOpen
  \bibfield  {author} {\bibinfo {author} {\bibfnamefont {J.~L.}\ \bibnamefont
  {Synge}},\ }\href@noop {} {\emph {\bibinfo {title} {Geometrical optics, an
  introduction to Hamiltons method}}}\ (\bibinfo  {publisher} {Cambrige
  University Press},\ \bibinfo {year} {1937})\BibitemShut {NoStop}%
\bibitem [{\citenamefont {Born}\ and\ \citenamefont {Wolf}(1999)}]{born99}%
  \BibitemOpen
  \bibfield  {author} {\bibinfo {author} {\bibfnamefont {M.}~\bibnamefont
  {Born}}\ and\ \bibinfo {author} {\bibfnamefont {E.}~\bibnamefont {Wolf}},\
  }\href@noop {} {\emph {\bibinfo {title} {Principles of optics, 7th ed.}}}\
  (\bibinfo  {publisher} {Cambridge University Press},\ \bibinfo {year}
  {1999})\BibitemShut {NoStop}%
\bibitem [{\citenamefont {van~de Hulst}(1980)}]{vandehulst80m}%
  \BibitemOpen
  \bibfield  {author} {\bibinfo {author} {\bibfnamefont {H.}~\bibnamefont
  {van~de Hulst}},\ }\href@noop {} {\emph {\bibinfo {title} {Multiple light
  scattering}}},\ Vol.\ \bibinfo {volume} {1 and 2}\ (\bibinfo  {publisher}
  {Academic Press},\ \bibinfo {year} {1980})\BibitemShut {NoStop}%
\bibitem [{\citenamefont {Ren}\ \emph {et~al.}(2011)\citenamefont {Ren},
  \citenamefont {Onofri}, \citenamefont {Roz\'{e}},\ and\ \citenamefont
  {Girasole}}]{RenVCRMOptLett2011}%
  \BibitemOpen
  \bibfield  {author} {\bibinfo {author} {\bibfnamefont {K.~F.}\ \bibnamefont
  {Ren}}, \bibinfo {author} {\bibfnamefont {F.}~\bibnamefont {Onofri}},
  \bibinfo {author} {\bibfnamefont {C.}~\bibnamefont {Roz\'{e}}},\ and\
  \bibinfo {author} {\bibfnamefont {T.}~\bibnamefont {Girasole}},\ }\bibfield
  {title} {\bibinfo {title} {Vectorial complex ray model and application to
  two-dimensional scattering of plane wave by a spheroidal particle},\
  }\href@noop {} {\bibfield  {journal} {\bibinfo  {journal} {Opt. Lett.}\
  }\textbf {\bibinfo {volume} {36(3)}},\ \bibinfo {pages} {370} (\bibinfo
  {year} {2011})}\BibitemShut {NoStop}%
\bibitem [{\citenamefont {Onofri}\ \emph {et~al.}(2015)\citenamefont {Onofri},
  \citenamefont {Ren}, \citenamefont {Sentis}, \citenamefont {Gaubert},\ and\
  \citenamefont {Pelc{\'e}}}]{Fabrice15OE}%
  \BibitemOpen
  \bibfield  {author} {\bibinfo {author} {\bibfnamefont {F.~R.}\ \bibnamefont
  {Onofri}}, \bibinfo {author} {\bibfnamefont {K.~F.}\ \bibnamefont {Ren}},
  \bibinfo {author} {\bibfnamefont {M.}~\bibnamefont {Sentis}}, \bibinfo
  {author} {\bibfnamefont {Q.}~\bibnamefont {Gaubert}},\ and\ \bibinfo {author}
  {\bibfnamefont {C.}~\bibnamefont {Pelc{\'e}}},\ }\bibfield  {title} {\bibinfo
  {title} {Experimental validation of the vectorial complex ray model on the
  inter-caustics scattering of oblate droplets},\ }\href
  {https://doi.org/10.1364/OE.23.015768} {\bibfield  {journal} {\bibinfo
  {journal} {Opt. Express}\ }\textbf {\bibinfo {volume} {23}},\ \bibinfo
  {pages} {15768} (\bibinfo {year} {2015})}\BibitemShut {NoStop}%
\bibitem [{\citenamefont {Yang}\ \emph {et~al.}(2015)\citenamefont {Yang},
  \citenamefont {Wu}, \citenamefont {Sheng},\ and\ \citenamefont
  {Ren}}]{YangCompJQSRT15}%
  \BibitemOpen
  \bibfield  {author} {\bibinfo {author} {\bibfnamefont {M.}~\bibnamefont
  {Yang}}, \bibinfo {author} {\bibfnamefont {Y.}~\bibnamefont {Wu}}, \bibinfo
  {author} {\bibfnamefont {X.}~\bibnamefont {Sheng}},\ and\ \bibinfo {author}
  {\bibfnamefont {K.~F.}\ \bibnamefont {Ren}},\ }\bibfield  {title} {\bibinfo
  {title} {Comparison of scattering diagrams of largen on-spherical particles
  calculated by {VCRM} and {MLFMA}},\ }\href@noop {} {\bibfield  {journal}
  {\bibinfo  {journal} {J. Quant. Spect. Rad.Trans.}\ }\textbf {\bibinfo
  {volume} {156}},\ \bibinfo {pages} {88} (\bibinfo {year} {2015})}\BibitemShut
  {NoStop}%
\bibitem [{\citenamefont {Duan}\ \emph {et~al.}(2021)\citenamefont {Duan},
  \citenamefont {Onofri}, \citenamefont {Han},\ and\ \citenamefont
  {Ren}}]{Duan2021OL}%
  \BibitemOpen
  \bibfield  {author} {\bibinfo {author} {\bibfnamefont {Q.}~\bibnamefont
  {Duan}}, \bibinfo {author} {\bibfnamefont {F.~R.~A.}\ \bibnamefont {Onofri}},
  \bibinfo {author} {\bibfnamefont {X.}~\bibnamefont {Han}},\ and\ \bibinfo
  {author} {\bibfnamefont {K.~F.}\ \bibnamefont {Ren}},\ }\bibfield  {title}
  {\bibinfo {title} {Generalized rainbow patterns of oblate drops simulated by
  a ray model in three dimensions},\ }\href {https://doi.org/10.1364/OL.434149}
  {\bibfield  {journal} {\bibinfo  {journal} {Opt. Lett.}\ }\textbf {\bibinfo
  {volume} {46}},\ \bibinfo {pages} {4585} (\bibinfo {year}
  {2021})}\BibitemShut {NoStop}%
\bibitem [{\citenamefont {Boyer}(1987)}]{Boyer1987}%
  \BibitemOpen
  \bibfield  {author} {\bibinfo {author} {\bibfnamefont {C.~B.}\ \bibnamefont
  {Boyer}},\ }\href@noop {} {\emph {\bibinfo {title} {The Rainbow, From Myth to
  Mathematics}}}\ (\bibinfo  {publisher} {Princeton University Press},\
  \bibinfo {year} {1987})\BibitemShut {NoStop}%
\bibitem [{\citenamefont {Adam}(2002)}]{Adam2002Rainbow}%
  \BibitemOpen
  \bibfield  {author} {\bibinfo {author} {\bibfnamefont {J.~A.}\ \bibnamefont
  {Adam}},\ }\bibfield  {title} {\bibinfo {title} {The mathematical physics of
  rainbows and glories},\ }\href@noop {} {\bibfield  {journal} {\bibinfo
  {journal} {Physics Reports}\ }\textbf {\bibinfo {volume} {356}},\ \bibinfo
  {pages} {229} (\bibinfo {year} {2002})}\BibitemShut {NoStop}%
\bibitem [{\citenamefont {Boitel}(1888)}]{Boitel1888}%
  \BibitemOpen
  \bibfield  {author} {\bibinfo {author} {\bibfnamefont {M.}~\bibnamefont
  {Boitel}},\ }\bibfield  {title} {\bibinfo {title} {Sur les arcs
  surnuméraires qui accompagnent l'arc-en-ciel},\ }\href@noop {} {\bibfield
  {journal} {\bibinfo  {journal} {Comptes rendus de l'Acad. Sci..}\ }\textbf
  {\bibinfo {volume} {CVI}},\ \bibinfo {pages} {1522} (\bibinfo {year}
  {1888})}\BibitemShut {NoStop}%
\bibitem [{\citenamefont {Lee}(1998)}]{LeeAO98}%
  \BibitemOpen
  \bibfield  {author} {\bibinfo {author} {\bibfnamefont {R.~L.}\ \bibnamefont
  {Lee}},\ }\bibfield  {title} {\bibinfo {title} {Mie theory, {A}iry theory,
  and the natural rainbow},\ }\href@noop {} {\bibfield  {journal} {\bibinfo
  {journal} {Appl. Opt.}\ }\textbf {\bibinfo {volume} {37}},\ \bibinfo {pages}
  {1506} (\bibinfo {year} {1998})}\BibitemShut {NoStop}%
\bibitem [{\citenamefont {Zhang}\ \emph {et~al.}(2022)\citenamefont {Zhang},
  \citenamefont {Rozé},\ and\ \citenamefont {Ren}}]{Zhang2022}%
  \BibitemOpen
  \bibfield  {author} {\bibinfo {author} {\bibfnamefont {C.}~\bibnamefont
  {Zhang}}, \bibinfo {author} {\bibfnamefont {C.}~\bibnamefont {Rozé}},\ and\
  \bibinfo {author} {\bibfnamefont {K.~F.}\ \bibnamefont {Ren}},\ }\bibfield
  {title} {\bibinfo {title} {Airy theory revisited with the method combining
  vectorial complex ray model and physical optics},\ }\href@noop {} {\bibfield
  {journal} {\bibinfo  {journal} {Opt. Lett.}\ }\textbf {\bibinfo {volume}
  {47}},\ \bibinfo {pages} {2149} (\bibinfo {year} {2022})}\BibitemShut
  {NoStop}%
\bibitem [{\citenamefont {Hovenac}\ and\ \citenamefont
  {Lock}(1992)}]{Hovenac92}%
  \BibitemOpen
  \bibfield  {author} {\bibinfo {author} {\bibfnamefont {E.~A.}\ \bibnamefont
  {Hovenac}}\ and\ \bibinfo {author} {\bibfnamefont {J.~A.}\ \bibnamefont
  {Lock}},\ }\bibfield  {title} {\bibinfo {title} {Assessing the contributions
  of surface waves and complex rays to far-field scattering by use of the
  {D}ebye series},\ }\href@noop {} {\bibfield  {journal} {\bibinfo  {journal}
  {J. Opt. Soc. Am. A}\ }\textbf {\bibinfo {volume} {9}},\ \bibinfo {pages}
  {781} (\bibinfo {year} {1992})}\BibitemShut {NoStop}%
\bibitem [{\citenamefont {Hovenac}\ and\ \citenamefont
  {Lock}(1993)}]{hovenac93}%
  \BibitemOpen
  \bibfield  {author} {\bibinfo {author} {\bibfnamefont {E.}~\bibnamefont
  {Hovenac}}\ and\ \bibinfo {author} {\bibfnamefont {J.}~\bibnamefont {Lock}},\
  }\bibfield  {title} {\bibinfo {title} {Calibration of the forward-scattering
  spectrometer probe: modelling scatteringfrom a multimode laser beam},\
  }\href@noop {} {\bibfield  {journal} {\bibinfo  {journal} {J. of Atmospheric
  and Oceanic Technology}\ }\textbf {\bibinfo {volume} {10}},\ \bibinfo {pages}
  {518} (\bibinfo {year} {1993})}\BibitemShut {NoStop}%
\bibitem [{\citenamefont {Duan}\ \emph {et~al.}(2023)\citenamefont {Duan},
  \citenamefont {Onofri}, \citenamefont {Han},\ and\ \citenamefont
  {Ren}}]{Duan2023OE}%
  \BibitemOpen
  \bibfield  {author} {\bibinfo {author} {\bibfnamefont {Q.}~\bibnamefont
  {Duan}}, \bibinfo {author} {\bibfnamefont {F.~R.~A.}\ \bibnamefont {Onofri}},
  \bibinfo {author} {\bibfnamefont {X.}~\bibnamefont {Han}},\ and\ \bibinfo
  {author} {\bibfnamefont {K.~F.}\ \bibnamefont {Ren}},\ }\bibfield  {title}
  {\bibinfo {title} {Numerical implementation of three-dimensional vectorial
  complex ray model and application to rainbow scattering of spheroidal
  drops},\ }\href@noop {} {\bibfield  {journal} {\bibinfo  {journal} {Optics
  Express}\ }\textbf {\bibinfo {volume} {31}},\ \bibinfo {pages} {34980}
  (\bibinfo {year} {2023})}\BibitemShut {NoStop}%
\bibitem [{\citenamefont {Deschamps}(1972)}]{Deschamps72}%
  \BibitemOpen
  \bibfield  {author} {\bibinfo {author} {\bibfnamefont {G.~A.}\ \bibnamefont
  {Deschamps}},\ }\bibfield  {title} {\bibinfo {title} {Ray techniques in
  electromagnetics},\ }\href@noop {} {\bibfield  {journal} {\bibinfo  {journal}
  {Proc. IEEE}\ }\textbf {\bibinfo {volume} {60(9)}},\ \bibinfo {pages} {1022}
  (\bibinfo {year} {1972})}\BibitemShut {NoStop}%
\bibitem [{\citenamefont {James}(1980)}]{JamesGTD80}%
  \BibitemOpen
  \bibfield  {author} {\bibinfo {author} {\bibfnamefont {G.~L.}\ \bibnamefont
  {James}},\ }\href@noop {} {\emph {\bibinfo {title} {Geometrical theory of
  diffraction for electromagnetic waves}}},\ \bibinfo {edition} {{IEEE}}\ ed.\
  (\bibinfo  {publisher} {Petter Peregrinus LTD.},\ \bibinfo {year}
  {1980})\BibitemShut {NoStop}%
\end{thebibliography}%

\end{document}